# Neuromodulators in food ingredients: insights from network pharmacological evaluation of Ayurvedic herbs

Neha Choudhary and Vikram Singh[*]

Centre for Computational Biology and Bioinformatics, School of Life Sciences, Central University of Himachal Pradesh, Dharamshala, India -- 176206.

[*]vikramsingh@cuhimachal.ac.in

**Abstract:**

The global burden of neurological diseases, the second leading cause of death after heart diseases constitutes one of the major challenges of modern medicine. Ayurveda, the traditional Indian medicinal systemenrooted in the Vedic literature and considered as a schema for the holistic management of health, characterizes various neurological diseases disorders (NDDs) and prescribes several herbs, formulations and bio-cleansing regimes for their care and cure. In this work, we examined neuro-phytoregulatory potential of 34,472 phytochemicals among 3,038 herbs (including their varieties) mentioned in Ayurveda using network pharmacology approach and found that 45% of these Ayurvedic phytochemicals (APCs) have regulatory associations with 1,643 approved protein targets. Metabolite interconversion enzymes and protein modifying enzymes were found to be the major target classes of APCs against NDDs. The study further suggests that the actions of Ayurvedic herbs in managing NDDs were majorly *via* regulating signalling processes, like, G-protein signaling, acetylcholine signaling, chemokine signaling pathway and GnRH signaling. A high confidence network specific to 219 pharmaceutically relevant neuro-phytoregulators (NPRs) from 1,197 Ayurvedic herbs against 102 approved protein-targets involved in NDDs was developed and analyzed for gaining mechanistic insights. The key protein targets of NPRs to elicit their neuro-regulatory effect were highlighted as CYP and TRPA, while estradiol and melatonin were identified as the NPRs with high multi-targeting ability. 32 herbs enriched in NPRs were identified that include some of the well-known Ayurvedic neurological recommendations, like, *Papaver somniferum*, *Glycyrrhiza glabra*, *Citrus aurantium*, *Cannabis sativa* etc. Herbs enriched in NPRs may be used as a chemical source library for drug-discovery agaisnt NDDs from systems medicine perspectives.

## 1 Introduction

Neurological disorders are amongst the most alarming health problems of mankind, responsible for approx. 9.1 million deaths worldwide. Neurological disorders were ranked as the leading cause of disabilitywith approx. 251 million cases following an increase of 7.4% between 1990 and 2015. Alzheimer's disease including other dementias are the most prevalent classes of neurological disorders accounting for approx. 40-50 million cases (Feigin et al., 2017). The complex behaviour of neurological diseases and disorders (NDDs) leads to various types of psychological, social and economic consequences. The social stigma for various neurological diseases is one of the reasons that undermine the treatment strategy. Although major advancements in biological understanding of neurological diseases and its treatment strategies have been carried out in past, the new generation drugs stillface the problem of psychiatric complications (Stroup and Gray, 2018). The use of synthetic drugs for brain disorders often follows an expensive and symptomatic long treatment;therefore, in such cases prescription of herbal medicines which are the most popular forms of CAM (complementary and alternative medicines) may be appreciated.

Ayurveda, the traditional medicinal system of India is not only a herb based disease management system, but a scientific approach to healthy living that includes fundamental principles of treatment and diagnosis. Ayurveda describes three forces that govern all the functional and structural processes of the human body. These include *Vata*, *Pitta* and *Kapha* often called *Tridosha*s or *body humors* described as biological energies found throughout the human body and mind. These *doshas* are based on 5 elements of nature & their associated properties where *Vata* refers to the qualities reflecting the elements of "space and air", *Pitta* "fire and water" and *Kapha* "water and earth". Nervous system disorders or neurological problems are mainly described under *Vatavyadhi*, the one caused by imbalance of *Vata*. The treatment procedure focuses on balancing the *Vata* and bringing harmony among the three elements. This is achieved *via* either herbal and herb-mineral based Ayurvedic preparations or *panchkarma* (bio-cleansing regimes) therapy (Mishra et al., 2013). The traditional literature is plentiful for the theories and paradigms for managing neurological and psychological disorders (Balsavar and Deshpande, 2014; Shamasundar, 2008). Ayurveda follows a holistic approach to treatment, describing a major role of nootropic herbs, both as a single herb or formulation-based with multi-dimensional bioactivity towards NDDs (Sharma et al., 2018). In the modern world also, Indian psychiatrists are open for the application of Ayurvedic concepts in their practice (Gupta, 1977).

Ayurveda describes food to possess three cosmic qualities, namely, Sattva (balance), Rajas (agitation), and Tamas (resistance), and consequently for mental wellbeing emphasizes on Sattvik diet that is mainly comprised of fresh fruits and vegetables, and most of the whole grains, legumes, nuts, and dairy products (Frawley, 2005). Several experimental studies have also shown the effect of natural compounds as neuromodulators and their implications in brain-related diseases eg., the treatment of cerebral injury using traditional medicinal herbs is attributed to the synergistic and multitargeting actions of their constituent phytochemicals (Wu et al., 2010); polyphenols, like, quercetin, catechin have been shown to possess a protective role in various animal models of neurological disorders (Heo and Lee, 2005; Mandel and Youdim, 2004).

In last couple of decades, the approach of drug design has been confined to the "single compound – single target – single pathway" approach. Given the advantage of multi-drug approach to deal drug resistance and treatment of complex diseases, multi-target pharmacology has fascinated the scientific community (Talevi, 2015). Network pharmacology (NP), a distinctive newer concept towards drug-discovery has gathered considerable attention in recent years (Hopkins, 2008). It follows apoly-pharmacological approachutilising the "multiple compounds – multiple targets" schema towards the disease treatment and has been applied to a great extent in providing scientific outlook to the traditional medicinal system (Choudhary et al., 2020; Choudhary and Singh, 2019; Li and Zhang, 2013). It is a way towards the new treatment strategies for various complex diseases, where conventional medical approaches fail to provide satisfactory results.

In this study, we utilised the network pharmacological approach towards examining the neuroregulatory aspects of the herbs mentioned in the Ayurveda. The methodology incorporates data collection at the levels Ayurvedic herbs, phytochemical identification of herbs, protein targets mapping of phytochemicals and NDDs-association of protein targets for their network construction and analyses. We believe that the study will provide a way towards the assembly of ancient Indian traditional medical knowledge, (i) to generate a repository of drug-like molecules from a natural source that can be used against NDDs, (ii) and to make recommendations for incorporating herbs in the diet chart for promoting the mental health.

## 2  Material and Methods:

### 2.1 Dataset of Ayurvedic herbs:

Indian medicinal plants database (IMPD) (http://www.medicinalplants.in/) is a repository of information about 7,258 Ayurvedic herbs used in Indian traditional medicinal system. To prepare a list of the Ayurvedic herbs, we made use of information available in IMPD, as on March, 2018. The herbs used in this study with their scientific names are available in **Supplementary Table-1.**

### 2.2 Phytochemical dataset of Ayurvedic herbs:

An extensive list of phytochemicals of each Ayurvedic herb was obtainedfrom the following sources, namely, IMPPAT (Indian Medicinal Plants, Phytochemistry And Therapeutics) (Mohanraj et al., 2018), TCM-MeSH (Zhang et al., 2017), CMAUP (CMAUP (Collective Molecular Activities of Useful Plants) (Zeng et al., 2019), PCIDB (PhytoChemical Interactions DB) (https://www.genome.jp/db/pcidb), NPASS (Natural Product Activity and Species Source database) (Zeng et al., 2018) and Duke's phytochemical database (https://phytochem.nal.usda.gov/phytochem/search).

Based on our earlier work (**Choudhary & Singh, 2021**) on 7,258 botanical names of Ayurvedic herbs mentioned in the IMPD, and their constituent 34,472 APCs from the aforementioned six databases, the current study is focused. The association of constituent APCs with each herb is provided in **Supplementary Table-1.**

### 2.3 ADMET descriptors estimation

To calculate the ADMET (Absorption, Distribution, Metabolism, Excretion and Toxicity) parameters of APCs, pkCSM pharmacokinetics server was used. This server, in addition of providing robust approachfor determining the pharmacokinetics and toxicity of small molecules, has been extensively validated with experimental data (Pires et al., 2015).

Absorption of the drug is an important pharmacokinetic parameterrepresenting the extent or availability of molecule or its active moiety to the circulatory system. Because of ease of administration and convenience for the patient, majority of the drugs are administered orally and hence good oral absorption is a prime property to optimize in new drug development. The human colon epithelial cancer cell line, Caco-2, is an important parameter for predicting the absorption of orally-administered drugs (Hubatsch et al., 2007). The cut-off values of Caco-2 permeability score of >0.9 and intestinal absorption value >30% were chosen to screen compounds with good absorption property (Choudhary and Singh, 2019). The ability of drug candidates to crossBlood-brain-barrier (BBB) is highly crucial, especially in the drug-design of neurotherapeutics. Threshold score of BBB value >-0.3 was also applied to the dataset for selecting compounds likely to cross BBB. Lastly, the toxicity profile of the compounds was checked and only the compounds negative for AMES and hepatotoxicity were chosen.

## 2.4 Protein Target identification of phytochemicals

Human proteintargets of APCs was obtained from STICH v5.0 (Szklarczyk et al., 2016), BindingDB (T. Liu et al., 2007) and PubChem (Bolton et al., 2008). STITCH utilises both manually curated as well as experimental data for cataloging chemical-target pairs (Szklarczyk et al., 2016). The high confidence interaction pairs with STITCH score of $\geq 0.7$ was used for data compilation. BindingDBis a publicly accessible web-platform contains experimental data of small molecule-protein binding interactions (Tiqing Liu et al., 2007). The targets from BindingDB were compiled using the chemical similarity value of $\geq 0.85$. From the chemical database of PubChem, the data of compounds with activity described as "Active" were collected for the analysis.

## 2.5 Disease association of the protein targets

DisGeNET, a publicly available discovery platform containing gene-disease associations linked to *Homo sapiens*was used to compile the information of various disease classes linked to the protein target set of APCs(Piñero et al., 2017). The database classifies disease data as per MeSH hierarchy and since, the focus of study relies on neuro-regulation, screening of disease-data was focused on nervous system and psychiatry related diseases. For this, disease-data of only those proteins that either correspond to C10 (Nervous system Diseases), all F branches (F0; Behavior and Behavior Mechanisms, F02; Psychological Phenomena and Processes, F03, Mental Disorders) or proteins categorized under "Mental or behavioral Dysfunction" were included for the analysis.

The proteins which come from "Mental or behavioural Dysfunction" class with no association from the four MeSH classes considered in this study comprises the "M" class, thus leading to the 5 types of disease classes i.e., C10, F0, F1, F2 and M.

## 2.6 Enrichment analysis

PANTHER v14.0 (Protein Analysis THrough Evolutionary Relationships) Classification System was used for protein classification and their enrichment analysis (Mi et al., 2019).

## 2.7 Network construction and analysis

Cytoscape v3.7 was utilized for construction and analysis of networks generated in this study. Protein-protein interaction (PPI) network was constructed using the data from STRING database v10.0 (Szklarczyk et al., 2015). Only high confidence interaction pairs with score ≥900 were used for the analysis. Module detection among the PPI network was carried out using MCODE (Molecular Complex Detection), a plug-in of Cytoscape (Bader and Hogue, 2003).

## Results and Discussion:

## 1.1 Herb-Phytochemical Network of Ayurvedic herbs (AH-APC network)

Based on 7,258 botanical names of AHs mentioned in the IMPD and their compiled 34,472 APCsan AH-APC (Ayurvedic herb – Ayurvedic phytochemicals) network was constructed (**Figure 1**). The network of AH-APC comprises of 133,203 edges among 37,510 nodes that include 3,038 herbs and their 34,472 APCs. C_06075 (Beta-sitosterol), C_00244 (Palmitic acid) and C_10144 (Quercetin) were the highly abundant APCs amongst the AHs, showing their presence in 618, 614 and 543 herbs, respectively. C_06075 is an important component of herbal products with high-therapeutic relevance especially in cardiovascular complications (Bin Sayeed et al., 2016), while C_10144 possess the ability to prevent and treat neurodegenerative diseases (Richetti et al., 2011). C_00244, represent palmitic acid and is the most abundant fatty acid present in human body which can be synthesised endogenously or provided in diet. A strict homeostasis of the molecule is maintained by the body responsible for governing various fundamental role among different biological processes (Carta et al., 2017). An abundance of phytochemicals of substantial therapeutic effects in AHs strengthens the rationale of applicability of traditional medicinal system in modern times.

**A.**

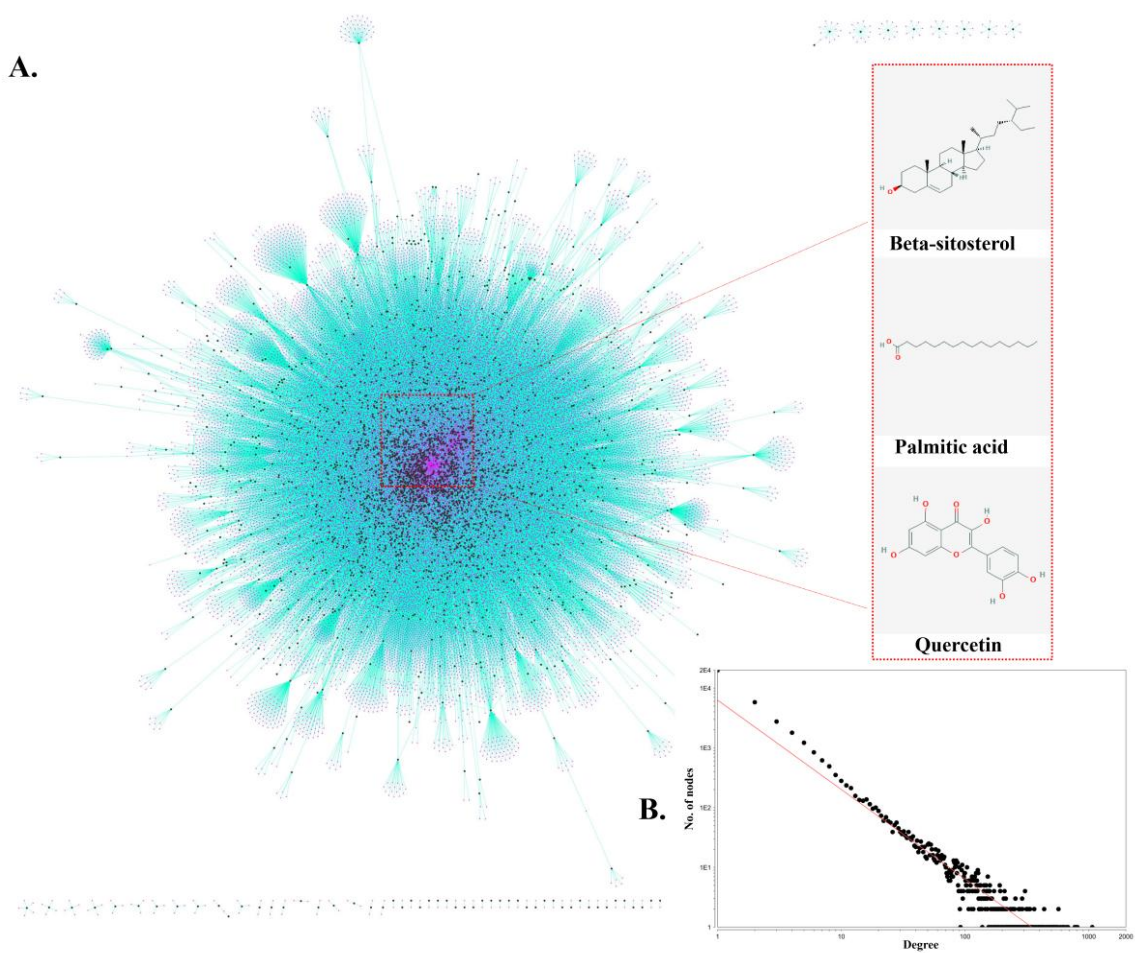

**Beta-sitosterol**

**Palmitic acid**

**Quercetin**

**B.**

**Figure 1: AH-APC Network**:

   A   **The AH-APC network.** It consists of 37,510 nodes and 13,203 edges representing associations of 34,472 APCs (pink coloured triangles) with 3,038 herbs (black circular nodes). The network is enriched with the phytochemicals, Beta sitosterol, Palmitic acid and Quercetin, showing its presence among 618, 614and 543 AHs.

   B   **Node-degree distribution of AH-APC Network.** Numbers of nodes are on x-axis, and their corresponding degree values (on a logarithmic scale) are represented on y-axis.

## 1.2 Phytochemicals - Protein target interaction network

As mentioned in Material and Methods section 2.2, each APC molecule was checked for its interactions with protein-targets, and all the obtained interactions were represented in the form of APC-PT (Ayurvedic PhytoChemical- Protein Target) network (**Supplementary Figure 1**). Satisfying the cut-off criterion used for protein-target compilation corresponding to high-confidence interaction pairs, only 18,004 of 34,472 APCs were found to possess protein-targeting ability against 8,443 human proteins, thus restricting the network size to 26,447 nodes (18,004 APCs+8,443 proteins) and 263,724 edges. The information of each APC-PT pair and its database source is given in **Supplementary Table 2.** The Venn diagram represents the contribution of each database source in determining the protein-target information **(Figure 2A).**

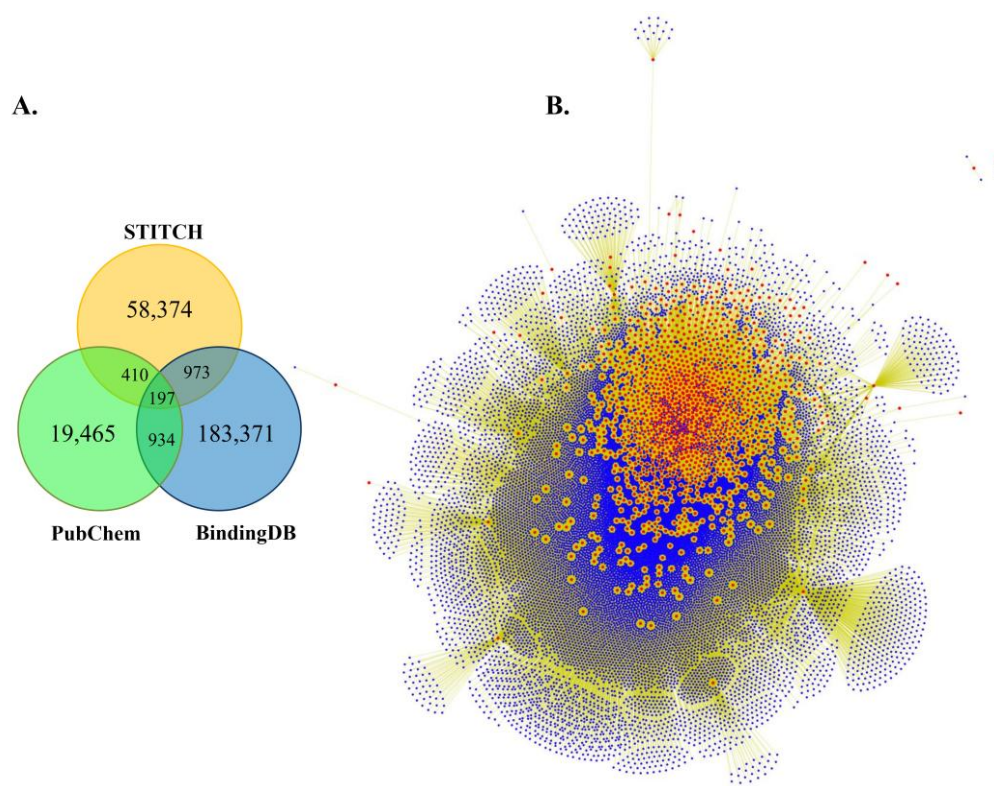

**Figure 2: Protein target pool of Ayurvedic herbs:**

A    Venn diagram showing the contribution of each protein-target (PT) prediction method in determining the targeting ability of APCs (Ayurvedic Phytochemicals). Each circle represents the contribution from target prediction algorithm and the numbers therein correspond to the APC-PT pairs being returned, resulting in a total of 263,724 unique pairs.

B    APC-APT neuroregulatory network: APC-APT (Ayurvedic Phytochemicals-Approved Protein Target) network showing the regulatory ability of 15,605 APCs to target 1,398 DrugBank approved protein target involved in neurological disease and disorders. The network is of size 17,003 nodes and 120,759 edges. The red coloured nodes represent APTs and blue coloured nodes are APCs.

When checked for the APCs capable of targeting an approved list of human proteins, as mentioned in DrugBank, 15,726 APCs were found to target 1,643 approved proteins. Ability of 45% of the APC pool of Ayurveda to regulate approved target proteins reflects the importance and ability of TIM to provide lead molecules for drug-discovery. The APC-APT (Ayurvedic Phytochemical - Approved Protein Target) pairs predicted by at-least two approaches used for target prediction, i.e., 2,515 pairs constitute the high-confidence pairs and denoted as "Dataset-I" (**Supplementary Table 3**) to be used in further analyses.

As already mentioned, only the protein-targets related to nervous system and its associated diseases were considered (Material and Methods section 2.5) which comprises of 4,894 of 8,443 PTs. This led to a subnetwork of size 17,003 nodes and 120,759 edges, referred to as APC-APT neuroregulatory network (Ayurvedic Phytochemical - Approved Protein Target) (**Supplementary Table 3**, **Figure 2B**), which comprises 1,398 PTs involved in NDDs (out of 4,894 among the approved target list in DrugBank). The network highlights the regulatory role of 15,605 APCs in APC-APT neuroregulatory network, of which 4,274 and 2,660 were found to target the hub proteins of the network, P18031 (encoded by PTPN1 gene; Tyrosine-protein phosphatase non-receptor type 1) and P00918 (encoded by CA2; Carbonic anhydrase 2). Inhibitors for both the proteins are well suited drug-candidates and constitutes the major targets for NDDs especially PTPN1 for Alzheimer's therapy and CA-inhibitors as anti-epileptic, with implication in cognitive impairment and phobias (Supuran, 2018; Vieira et al., 2017).

### 1.3 Protein-target-Disease Network

The neuro-regulatory effects of AHs are obtained by accessing the disease association profile of the protein targets of each constituent APC molecule. Satisfying the criteria, 4,894 of 8,443 protein targets show their association with 3,347 types of NDDs (based on DisGeNET disease IDs). The disease IDs were grouped into 5 classes, as mentioned in Material & Methods Section 2.5, thus leading to the disease-network of size 4,899 (4,894 proteins and 5 disease classes) and 11,789 edges (**PT-NDD network, Protein Target- Neurological Diseases and Disorder network, Figure 3A, Supplementary Table 4**). The complete disease-mapping of 4,894 protein targets with DisGeNET disease IDs, disease class and UMLs semantic class type is given in **Supplementary Table 4.**

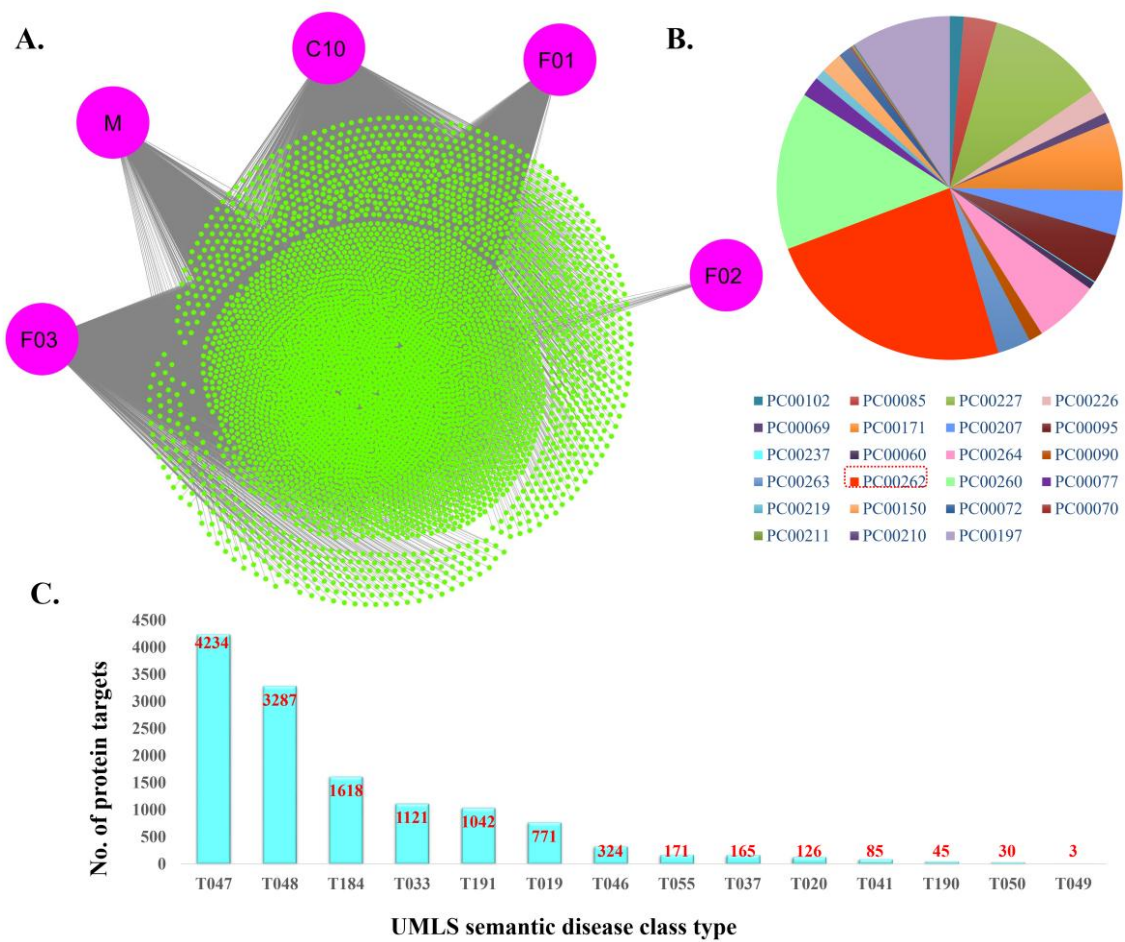

**Figure 3: PT-NDD (Protein Target- Neurological Diseases and Disorder) network**

A   PT-NDD network consisting of 4,894 protein-targets associated with NDDs leading to the disease-network of size 4,899 (4,895 proteins and 5 disease classes) and 11,789 edges. The pink circular nodes represent the 4 MeSH classes that include C10 (Nervous system Diseases), F01 (Behavior and Behavior Mechanisms), F02 (Psychological Phenomena and Processes) and F03 (Mental Disorders), and M represents the class of proteins which come from "Mental or Behavioral Dysfunction" class with no association from the four MeSH classes.

B   Protein-classification of 4,894 proteins of PT-NDD network showing the distribution among 23 PANTHER-Protein class, where metabolite interconversion enzymes (PC00262) and protein modifying enzymes (PC00260) were the major classes with 23.8% and 14.9% proteins.

C   The proteins of PT-NDD network and their grouping onto disease semantic type class provided by UMLS (Unified Medical Language System). The 14 UMLS semantic classes are T047 (Disease or Syndrome); T048 (Mental or Behavioral Dysfunction); T184 (Sign or Symptom); T033 (Finding); T191 (Neoplastic Process); T019 (Congenital Abnormality); T046 (Pathologic Function); T055 (Individual Behavior); T037 (Injury or Poisoning); T020 (Acquired Abnormality); T041 (Mental Process); T190 (Anatomical Abnormality); T050 (Experimental Model of Disease) and T049 (Cell or Molecular Dysfunction).

The proteins of the PT-NDD network were majorly from the class of metabolite interconversion enzymes (PC00262, 23.8%) and protein modifying enzymes (PC00260, 14.9%) (**Figure 3B**). The proteins belong to 23 protein-classes as identified using PANTHER and the protein-classification is given in **Supplementary Table 5.**

The proteins were grouped according to the disease semantic type class provided by UMLS (Unified Medical Language System) and each class was studied in detail for their network composition and properties. The classes include T047 (Disease or Syndrome); T048 (Mental or Behavioral Dysfunction); T184 (Sign or Symptom); T033 (Finding); T191 (Neoplastic Process); T019 (Congenital Abnormality); T046 (Pathologic Function); T055 (Individual Behavior); T037 (Injury or Poisoning); T020 (Acquired Abnormality); T041 (Mental Process); T190 (Anatomical Abnormality); T050 (Experimental Model of Disease) and T049 (Cell or Molecular Dysfunction). The distribution of protein targets among each UMLS class is given in (**Figure 3C**). The figure helps to find that the protein targets of APCs affect the nervous system *via* their involvement in various types of abnormalities, neoplastic, mental dysfunctions. Majorly the APCs have effects on T047 and T048 with their protein targeting ability of 86 % and 67%, respectively. To further ascertain the association of these top UMLS target classes in the overall neuro-regulatory network of humans, the PPI analysis was conducted.

**Case study I: Construction and analysis of UMLS class-based PPI networks**

According to 4,894 protein targets in PT-NDD network, the PPI networks were constructed *via* mapping the proteins onto high-confidence human-PPI. The high confidence network was obtained by STRING using confidence score ≥900.

A sub-PPI network specific to T048 and T047 was constructed and analysed in detail to check how the modular architecture of PPI pathologically characterises the processes associated with NDDs. For the analyses, only giant component of PPIs was considered and isolated nodes were removed. In this manner, sub-PPIs were constructed, and both T048-PPI and T047-PPI were subjected to module detection. The candidate proteins of each cluster were analyzed for gene-function analysis using pathway association.

The PPI specific to T048 were found to involve the association among 2,579 protein-targets, leading to the PPI size of 2,579 nodes and 21,987 edges with an average number of 17 neighbours. The regulatory hub of the PPI was found to be P04637 and P12931 with the ability to regulate 810 and 646 other mental dysfunction associated protein. Targeting action of 137 APCs towards P04637 and 445 towards P12931 reflects the central role of these APCs in controlling the whole PPI. Where P04637 is encoded by TP53 (tumor antigen p53) and is found to be responsible for neurocognitive effects (Papiol et al., 2004), P12931 (Proto-oncogene tyrosine-protein kinase) protein encoded by SRC gene with implications in schizophrenic (Banerjee et al., 2015) and Alzheimer's cases (Beirute-Herrera et al., 2020) and thus provide a mechanistic role of these APCs in managing mental and behavioural dysfunctions.

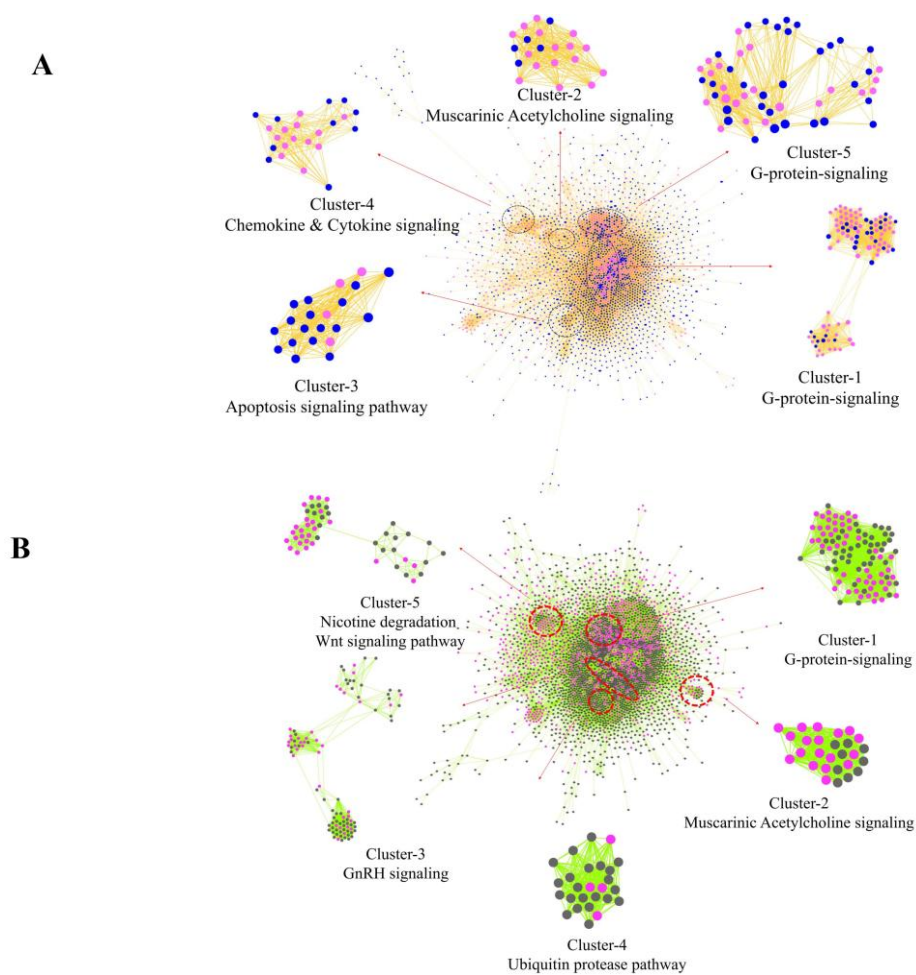

**Figure 4: Case-study 1: UMLS class specific PPI networks and their modular analysis**

A  PPI-network specific to T048 consisting of 2,579 nodes and 2,1987 edges showing the interaction among 2,579 proteins of T047 class. The PPI is subjected for cluster analysis, highlighting the role in multiple signalling processes. The pink circular among the PPI represents the location of approved protein targets of phytochemicals of AHs.

B  PPI-network specific to T047 consisting of 3,317 nodes and 30,769 showing the interaction among 2,579 proteins of T047 class. The clusters are arranged around the PPI and subjected for pathway enrichment, highlighting the regulatory role of protein targets of APCs mainly *via* controlling signalling processes. The pink circular nodes among the PPI represents the location of approved protein targets of phytochemicals in AHs.

The top-5 modules among the T048-PPI are represented in the **Figure 4A** and were found to be associated with multiple signaling processes, representing the true complex pathophysiology of NDDs and ability of Ayurveda in managing the same. Similarly, when checked for the T047 specific PPI of network size of 3,317 nodes and 30,769 edges, the mechanistic effect of AHs in NDDs were highlighted. As seen in **Figure 4B**, the top clusters have their involvement in the pathways associated with signaling i.e., G-protein, Acetylcholine, Wnt etc. The PPI data and the information of identified clusters with their protein-composition and functional enrichment is given in **Supplementary Table 6.**

The PPI analysis highlights the fact that APCs mainly work via controlling G-protein signaling, acetylcholine signaling, chemokine signaling pathway and GnRH signaling.

### Case study II: Schizophrenia specific regulatory network

Schizophrenia, a chronic disorder characterised by hallucinations is a severe mental-disorder affecting 20 million people across the globe (James et al., 2018). Hallucination in general terms refers to condition where the affected individual sees or hears something that does not exists. The classical Ayurvedic texts describe the condition as *mithyajnana* (false perceptions), *maya* (illusions), *moha* (infatuations), or *bhrama* (confusion) which comes under *unmade* (mental disorder). The terms represent the disordered state of mind, where individual loses its power to regulate its actions and conduct as per the rules of the society (Balsavar and Deshpande, 2014). The terms cover all the aspects of psychotic disorders as governed by modern scientific notion. To decipher in detail the herbs effective in alleviating schizophrenia, a tetra-partite network of Disease--Protein-target--Phytochemical--Herbs was constructed and analysed (**Figure 5**). When checked for the herbal composition in the network, it was interesting to note the regulatory action of 569 APCs towards 189 Schizophrenia associate proteins. The APC-PT associations come from the high confidence pool, *i.e.,* **Dataset-I.** The herbal association of the 569 APCs comprises of 1,900 AHs. Only AHs having more than 20 APCs targeting Schizophrenia proteins were used for the network construction. In this manner 91 AHs, their 381 APCs, their targeting ability against 169 Schizophrenia proteins in their involvement in 11 types of Schizophrenia (**Schizophrenia specific regulatory network; SSR network; Figure 5**) leading to network of size 652 nodes 3,581 edges. The major targets of APCs in the network are cytochrome P450 protein (P05177; CYP1A2 gene) and Prostaglandin G/H synthase 2 (P35354; PTGS2 gene) where polymorphism of P05177 is associated with dyskinesia (Basile et al., 2000) while of PTGS2 is linked to niacin skin flush response (Nadalin et al., 2013) in schizophrenic patients. The APCs that contribute maximally in targeting Schizophrenia-proteins in the SSR-network comes from AH_2070 (*Panax ginseng*) and AH_1360 (*Glycyrrhiza glabra*) with 54 and 44 APCs to target Schizophrenia proteins. The detail of complete SSR network and its associated 91 Ayurvedic herbs with more than 20 APCs targeting Schizophrenia proteins is detailed in **Supplementary Table 7.**

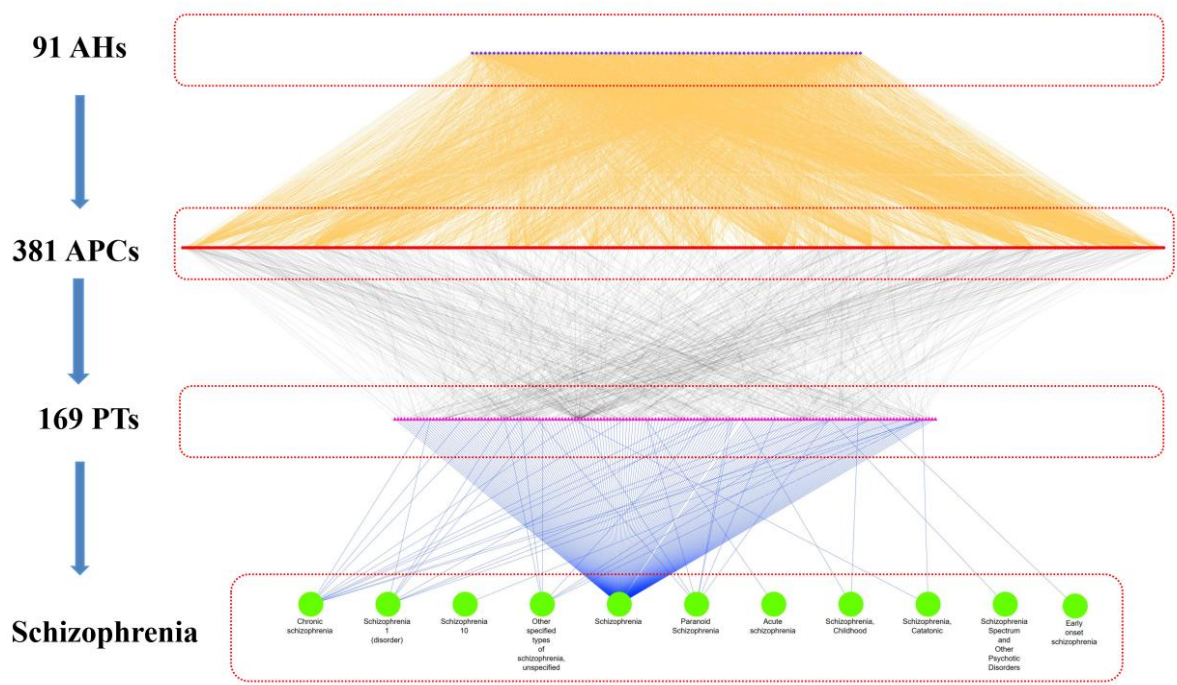

**Figure 5: Schizophrenia specific regulatory network (SSR network):** Tetrapartite SSR network highlighting the role of 91 AHs (I-layer) their 381 APCs (II-layer) in targeting 169 Schizo proteins (III-layer) involved in 11 types of Schizophrenia (IV-layer) leading to a network of size 652 nodes and 3,581 edges. For the representation, the association among each layer is coloured differently. Only AHs with more than 20 APCs targeting Schizophrenia associated proteins were used for the SSR network construction.

C_10144 (Quercetin, PubChem ID: 5280343), C_02666 (Glutamic acid, PubChem ID: 33032), C_00019 (gamma-Aminobutyric acid, PubChem ID: 119) were the APCs highlighted among the SSR-network for their high-multi-targeting ability by targeting 13, 12 & 12 Schizophrenia-proteins, respectively. Quercetin is already a well-known anti-psychotic agent (Dajas et al., 2015; Mert et al., 2019), whereas role of glutamic acid and gamma-Aminobutyric acid system in Schizophrenia and various mood disorders is well-studied (Howes et al., 2015; Wassef et al., 1999). This highlights the ability of the network approach in highlighting potential APCs and their targeting action on various NDDs. In the similar manner other types of NDDs can be focused and studied in detail for their targeting mechanism.

**Case study III:   Prioritising neuro-phytoregulatory Ayurvedic herbs**

To decipher in detail the herbs effective in alleviating NDDs, the complete network data was compiled to form a tetra-partite network of Disease--Protein-target--Phytochemical--Herbs. For achieving this, each layer was added in a step-by-step manner, where firstly the APCs layer was chosen. For this the APCs passing the pharmacokinetic criteria as discussed in Material & Methods section 1.3 were considered. For the selected APCs, only the proteins belonging to the approved-target class were considered. Only those interactions between APC-PT network that come from the high-confidence pool (Dataset-I) were opted for the network construction. For the screened-in APCs, the herb association was taken from AH-APC network, and the disease associations maintaining the threshold criteria (score>0.05) (Choudhary et al., 2020) in the PT-NDD network were considered. In this manner, a high confidence tetra-partite network consisting of 1,197 AHs, their 219 APCs, and their 102 approved class of target proteins involved in different types of NDDs was developed.

The putative bio-active 219 APCs passing the pharmacokinetic criteria are referred to as Neuro-phytoregulators (NPRs) and the resulting network as NPR-network (**Supplementary Table 8**). Among the list of 219 NPRs, the multi-targeting nature was checked and it was interesting to note that 76 possess targeting ability against >1 proteins, highest by C_00672 (PubChem ID:5757; Estradiol), C_00219 (PubChem ID: 89; Melatonin) towards 10 and 9 proteins while other 143 are specific to their protein targets. Estrogens play an important role in brain *via* various routes and exert a plethora of neuroprotective actions. Estradiol, a potent steroid involves MAPK and PI3k signaling to offer neuroprotection and have been shown to possess neuroprotective activity in variety of cellular and animal models (Petrovska et al., 2012). While melatonin possesses both neuroprotective and neural plasticity owing to its anti-oxidant, anti-inflammatory and anti-excitotoxicity activity in neurons (Lee et al., 2019). The major targets of NPR-network are cytochrome P450 protein (P05177; CYP1A2 gene) and Transient receptor potential cation channel subfamily A member 1 (O75762; TRPA1 gene), being targeted by 37 and 28 NPRs respectively. CYPs are the class of functionally active enzymes at the neurovascular interface and the altered expression of the same has been reported to be involved in various NDDs (Ghosh et al., 2016). While TRP proteins belong to a class of cation channel with implications in diverse physiological process of brain as well development of NDDs (Wang et al., 2020).

When checked for the AHs, whose more than 15 NPRs target NDDs associated proteins among the NPR-network, 32 herbs were highlighted. The information of all the 32 AHs is detailed in **Supplementary Table 9.** Among the list, herbs like AH_2085 (*Papaver somniferum*), AH_1360 (*Glycyrrhiza glabra*), AH_3034 (*Vitis vinifera*) and AH_3088 (*Zingiber officinale*), AH_0699 (*Citrus aurantium*), *AH_0949 (Daucus carota), AH_0526 (Cannabis sativa), AH_2219 (Piper nigrum) etc.* are the prominent Ayurvedic recommendations that have been traditionally used for treating NDDs or their associated effects and are highlighted in a study (Sharma et al., 2018) (**Figure 6**).

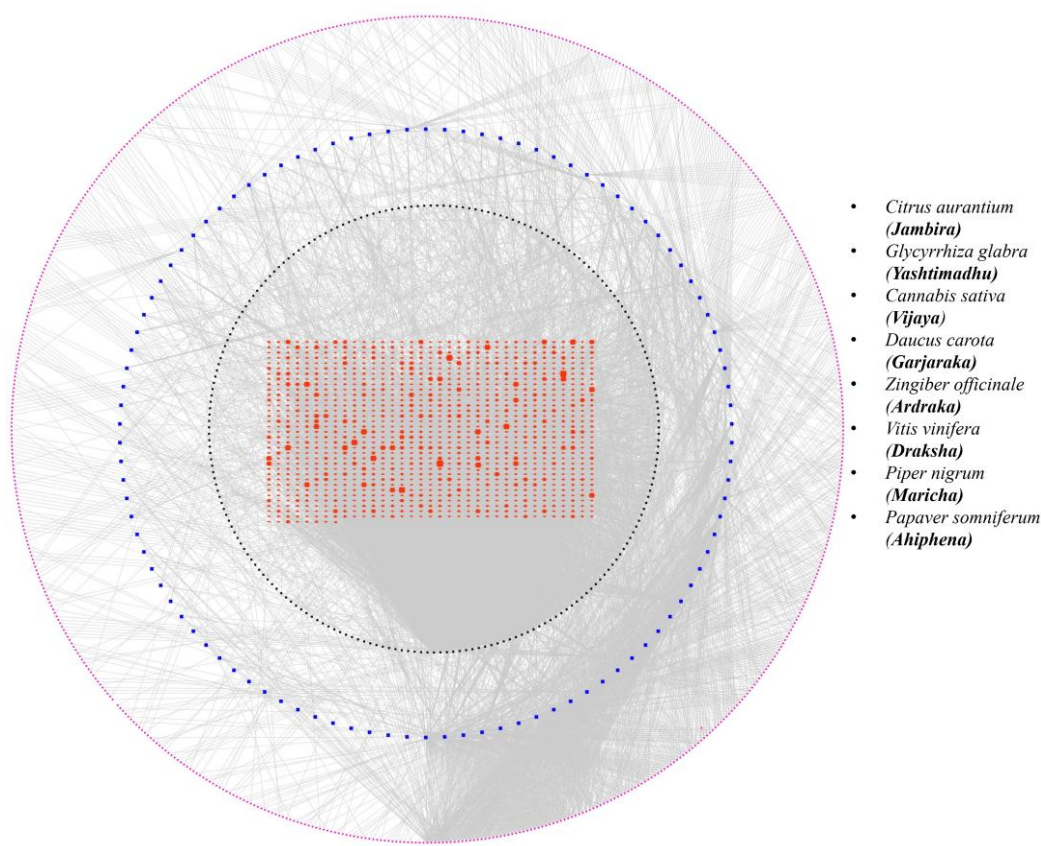

- *Citrus aurantium (Jambira)*
- *Glycyrrhiza glabra (Yashtimadhu)*
- *Cannabis sativa (Vijaya)*
- *Daucus carota (Garjaraka)*
- *Zingiber officinale (Ardraka)*
- *Vitis vinifera (Draksha)*
- *Piper nigrum (Maricha)*
- *Papaver somniferum (Ahiphena)*

**Figure 6: NPR (Neuro PhytoRegulatory) Network:** Four-layer NPR network showing the targeting profile of pharmaceutically relevant 219 NPRs (black coloured layer) distributed among 1197 herbs (centrally placed red circular nodes) towards 102 approved class of target proteins (blue coloured layer) involved in different types of NDDs (outer most pink layer) with disease association score >0.05. Of 32 herbs with high neuro-regulatory power identified, few well known Ayurvedic neurological recommendations are mentioned in the figure.

Thus, suggesting the efficacy of NP-approach in highlighting neurophytoregulatory herbs and their underlying phytochemical specific protein target role towards NDDs. A subnetwork of NPR specific to these 32 AHs can be checked in **Supplementary Figure-2** and the interactions in **Supplementary Table 9** for detailed investigations. The NPR-network can be searched for any particular interest of NDD and can be back-traced to search for potential regulatory NPRs and their source herb.

**Summary:**

With an upsurge in lifestyle disorders, the chances of NDDs are believed to increase at higher pace in near future. The pathogenesis and mechanism of development of NDDs are still not fully understood, thus posing a great challenge towards developing treatment regimens with greater efficacy and least toxicity. Herbal remedies are gaining great popularity across the globe due to their holistic approach to disease treatment and prevention.

Ayurveda, the traditional medicinal system is a healthy lifestyle practiced by the people of India for the past 5,000 years. Ayurvedic texts describe various treatment and preventive regimes for the well-being of mankind. Although the use of Ayurvedic system in various NDDs has already been suggested and is still followed by Indian psychiatrists, the lack of robust confirmation and evidence-based studies has limited their global recognition. Considering the aspect, the current study is envisaged for examining neuroregulatory aspect of Ayurvedic herbs and their contributing phytoconstituents *via* applying the Network pharmacological concept. Information of 7,258 botanical Ayurvedic herbs and their 34,472 phytochemicals form the basis of the study. Phytochemicals like C_06075 (Beta-sitosterol), C_10144 (Quercetin) were of high abundance among the Ayurvedic herbs. To understand the pharmacological effects of 34,472 APCs among human-system the protein target (PT) information was collected, identifying the role of 45% of APC-pool in regulating 1,643 approved classes of proteins. To limit the network studies specific to NDDs, only the protein targets involved in NDDs were focused. Of 8,443 PTs of Ayurvedic herbs, 4,894 show their participation in NDD pathogenesis and development. AHs were found to manage NDDs mainly via targeting proteins of metabolite interconversion enzymes and protein modifying enzymes. The modular architecture of NDD specific PPIs were also investigated and it found that APCs mainly work via controlling G-protein signaling, acetylcholine signaling, chemokine signaling pathway and GnRH signaling. As a case study, Schizophrenia is explored in detail showing the role of 91 AHs, especially, *Panax ginseng* and *Glycyrrhiza glabra* in managing the disease. Quercetin, Glutamic acid and gamma-Aminobutyric acid were amongst the top multi-targeting molecules towards Schizophrenia proteins.

A high confidence tetra partite network (NPR-network) was constructed in the study and is believed to be a substantial important resource for deciphering the component wise systemic level effects of Ayurvedic herbs in managing NDDs. The identification of putative bio-active 219 APCs (referred to as NPRs) identified in this study represents a class of future lead molecules that can be investigated for future *in-vitro* and *in-vivo* studies. Among these, estradiol and melatonin possess very high multitargeting ability. The NPRs mainly target CYP and TRPA proteins to elicit their therapeutic effect and comes from well-known Ayurvedic recommendations like *Papaver somniferum*, *Glycyrrhiza glabra*, *Zingiber officinale*, *Cannabis sativa, Piper nigrum, Citrus aurantium etc.* The detailed information of association data of these 219 NPRs, with their 1,197 herbal compositions and their 102 protein-target information given in this study can be used for identifying the multi-targeting phytoconstituents and their underlying phytochemical specific protein-targeting role against NDDs. This will be helpful in multiple aspects, which involves identification of key phytochemicals for quick development of new drug molecules, finding herb-specificity against a particular protein target. The repository of NPR-protein target pairs associated with NDDs, will be highly helpful to search for structural scaffolds responsible for their pharmacological action and their interaction behavior around the protein space. We believe that the steps followed in this study and the results generated will be of great help in understanding the role of Ayurvedic herbs in neuroprotection and management of various neurological diseases and disorders. The study may be considered as a giant leap towards integrating the modern scientific approaches with traditional knowledge to offer a scientific outlook to the traditional therapies.

**Acknowledgements:** N.C. is grateful to the Indian Council of Medical research (ICMR) for support provided through ICMR-SRF.

**Authors Contribution:** VS conceptualized the study and designed the research framework. NC contributed to compilation of data and performed computational experiments. NC and VS analyzed the results and prepared the manuscript.

**Data Availability:** All the data generated in this work is available for download at https://drive.google.com/file/d/1KTIugL4AgPexrvDYLYSCAjL4JISZL8g8/view?usp=sharing

**Conflict of interest:** Authors declare that there is no conflict of interest regarding the publication of this work.